\documentclass[preprint,12pt]{elsarticle}

\usepackage{graphicx}
\usepackage{longtable}
\usepackage{tabularx}
\usepackage{url}
\usepackage[width=135mm]{caption}

\usepackage[labelformat=simple]{subcaption}

\graphicspath{{./}}

\biboptions{square,numbers,sort&compress}
\usepackage{subfig}

\journal{Proceedings of the Combustion Institute}

\usepackage{amssymb,amsmath}
\usepackage[usenames]{color} 
\usepackage[version=3]{mhchem}

\def\one{88mm}

\newcommand{\ele}{\ce{e-}}
\newcommand{\hyd}{\ce{H3O+}}
\newcommand{\for}{\ce{HCO+}}
\newcommand{\iV}{\mathit{i\;\!\textnormal{-}\:\!\!V}}
\newcommand{\f}{\frac}

\begin{document}
\begin{frontmatter}

\title{The $\iV$ curve characteristics of burner-stabilized premixed flames:\\
detailed and reduced models} 

\author[add1]{Jie Han}
\author[add1]{Memdouh Belhi}
\author[add2]{Tiernan A. Casey}
\author[add3,add1]{Fabrizio Bisetti\corref{cor1}}
\cortext[cor1]{Corresponding author}
\ead{fbisetti@utexas.edu}
\author[add1]{Hong G.~Im}
\author[add2]{Jyh-Yuan~Chen}

\address[add1]{Clean Combustion Research Center, King Abdullah University of Science and Technology, Thuwal 23955, Saudi Arabia}
\address[add2]{Department of Mechanical Engineering, University of California at Berkeley, Berkeley, CA 94720-1740, USA}
\address[add3]{Department of Aerospace Engineering and Engineering Mechanics, University of Texas at Austin, Austin, TX 78712-1085, USA}

\begin{abstract}
The $\iV$ curve describes the current drawn from a flame as a
function of the voltage difference applied across the reaction zone.
Since combustion diagnostics and flame control strategies
based on electric fields depend on the amount of current
drawn from flames, there is significant interest in modeling
and understanding $\iV$ curves.
We implement and apply a detailed model
for the simulation of the production and transport of ions and electrons in
one-dimensional premixed flames.
An analytical reduced model is developed based on the detailed one, and analytical expressions are used
to gain insight into the characteristics of the $\iV$ curve for various flame configurations.
In order for the reduced model to capture the spatial distribution of the electric field accurately,
the concept of a \emph{dead zone} region, where voltage is constant, is introduced, and a
suitable closure for the spatial extent of the dead zone is proposed and validated.
The results from the reduced modeling framework are found to be in good agreement with those from the detailed simulations.
The saturation voltage is found to depend significantly on the flame location
relative to the electrodes, and on the sign of the voltage difference applied.
Furthermore, at sub-saturation conditions, the current is shown to increase linearly or quadratically with the applied voltage,
depending on the flame location.
These limiting behaviors exhibited by the reduced model elucidate the features of $\iV$ curves observed experimentally.
The reduced model relies on the existence of a thin layer where charges are produced, corresponding to the reaction
zone of a flame.
Consequently, the analytical model we propose is not limited to the study of premixed flames,
and may be applied easily to others configurations, e.g.~nonpremixed counterflow flames.
\end{abstract}


\end{frontmatter}

\section{Introduction}\label{sec:introduction}

\begin{figure}[h]
\centering
\includegraphics[width=\one]{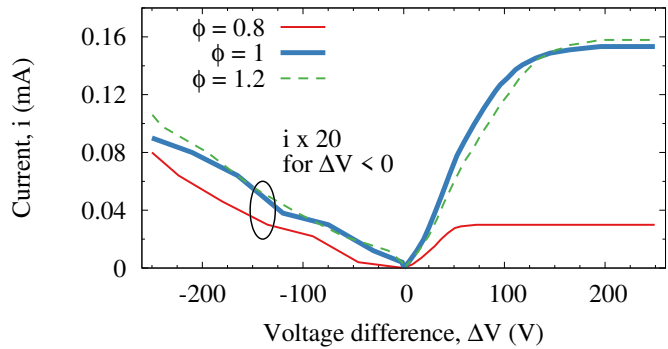}
\caption{Current $i$ versus the applied electric difference $\Delta V=V_R-V_L$
measured experimentally in burner-stabilized methane/air premixed flames with various
equivalence ratios at atmospheric pressure~\citep{speelman2015}.
The data is taken from Figure 13 in Ref. \citep{speelman2015}.
In the experimental setup, the right electrode at $V_R$ consists
of a metal mesh placed in the burnt gases 1 cm from the surface of a
grounded heat flux burner, which serves as the left electrode at $V_L$~\citep{speelman2015}.
The area of the burner's surface is $A=7$ cm$^2$.
The saturation currents for positive voltages are 0.03, 0.15, and 0.16 mA for
the lean, stoichiometric, and rich flames, respectively.
For the sake of clarity, the data for $\Delta V < 0$ are multiplied by 20.}
\label{fig:ivspeelman}
\end{figure}

When a voltage is applied across a hydrocarbon flame,
the current drawn from the flame increases with the voltage.
The relationship describing this behavior is known as an ``$\iV$ curve''.
Valuable information on the ionic structure of a flame 
can be obtained from a detailed understanding of $\iV$ curves \cite{electriclaspect1969}.
$\iV$ curves have been investigated in the combustion literature
with the aim of characterizing the production rates and
densities of charged species \cite{Goodings2001,ddrankiniV2015}
and analyzing electric fields effect on flames \cite{speelman2015,samedrivation2011}.
An example of the $\iV$ curve is shown in Fig.~\ref{fig:ivspeelman}
for a burner-stabilized premixed methane/air flame at atmospheric pressure~\cite{speelman2015}.
These curves have a strong dependence on the type of fuel, flame stoichiometry,
pressure, and position of the flame with respect to the electrodes.

The experimentally measured $\iV$ curve in Fig.~\ref{fig:ivspeelman} shows that the current
increases monotonically for positive voltages up to a threshold value of $\Delta V$,
whereby the current saturates and reaches a plateau.
Increasing the voltage beyond this point does not alter the current, so that
the \emph{saturation voltage} $\Delta V_s$ and the \emph{saturation current} $i_s$ are
identified.
Similar data is reported from experimental studies using
counterflow flame~\citep{yxiV2016} and coflow flames~\citep{ddrankiniV2015}.
The $\iV$ curves in Fig.~\ref{fig:ivspeelman} depend on the equivalence ratio of the flame
and show a significant degree of asymmetry, with the current increasing
at a much slower rate for negative potential differences than positive ones.

The $\iV$ characteristics of flames are a \emph{macroscopic} manifestation
of the complex interaction of chemically-driven generation of charges
and charge transport induced by the electric field due to the applied voltage.
Since it is difficult to determine experimentally the spatial distribution of the charges,
current, voltage, and electric field, numerical simulations are an ideal tool for investigating
the ionic structure of flames under applied voltages~\citep{jieiontransport2015,yxiV2016}.

In this paper, we aim to describe the mechanism controlling the current drawn
from a flame under varying applied voltages.
An analytical, reduced model explaining
the key features of the $\iV$ curve is proposed and its accuracy is assessed
against detailed numerical simulations of charges and voltage in 
burner-stabilized premixed methane/air flames.
Our analytical model for the quantitative characterization of $\iV$ curves may be of use in further developing combustion
diagnostics based on current, and strategies for controlling flames using
electrohydrodynamic forces~\citep{electriclaspect1969}.

\section{Configuration, models, and methods}\label{sec:methods}

The configuration consists of a one-dimensional, steady, burner-stabilized methane/air flame at atmospheric pressure.
The analysis is carried out for a stoichiometric mixture and selected simulations are repeated
for two lean flames ($\Phi = 0.8,0.9$) in order to assess the generality of the proposed analytical model. 
The temperature of the mixture at the inlet is 350 K and the mass flow rate for the stoichiometric flame is
0.036 g cm$^{-2}$ s$^{-1}$.
A domain size $L$ of 1 cm is chosen and, since the voltage is applied at the boundaries of the computational domain,
$L$ corresponds also to the distance between the electrodes.
The flame parameters and computational setup considered here are inspired by the experimental setup
for the characterization of the electrical response of burner-stabilized premixed flames in Ref.~\citep{speelman2015}.

The PREMIX~\citep{premix-chemkin} code was modified to allow for the simulation of the steady one-dimensional
spatial distribution of ions and electrons across a burner-stabilized flame under an applied voltage.
Firstly, the Poisson equation describing the voltage distribution was added to the nonlinear system
of equations.
Dirichlet boundary conditions for the voltage at the left and right domain boundaries were implemented.
Secondly, the mass flux of the charged species was modified to include the drift
flux~\citep{jieiontransport2015} in addition to the diffusive fluxes.
Lastly, the boundary conditions proposed in Ref.~\citep{DCACBelhi2013} were adopted for
the balance equations describing the transport of charged species.

The oxidation of methane is described by a skeletal methane/air mechanism~\citep{skelGRI}
to which \ce{CH2},~\ce{CH2(S)}, and~\ce{CH} and related reactions from GRI Mech 3.0~\citep{grimech30} were added.
The neutral mechanism assembled consists of 19 species and 64 reactions.
The thermodynamic, kinetics, and transport data for neutral species are taken from Ref.~\citep{grimech30}.
The formation of charged species, charge transfer, and charge recombination reactions are described
using a mechanism that includes the formyl cation \for, the hydronium ion \hyd,
and electron \ele,~and consisting of the following 4 reactions:
\ce{CH + O ->[k_1] HCO+ + e-} (R1);
\ce{HCO+ + H2O <-> H3O+ + CO} (R2);
and \ce{H3O+ + e- ->[k_{3a},k_{3b}] neutrals} (R3a, R3b). 
These are the chemi-ionization (R1), proton transfer (R2),
and recombination (R3) reactions with rate constants $k=AT^n\exp(-E/\mathcal{R}T)$,
where $A_1= 2.51\times10^{11}$, $n_1=0$, and $E_1=7.12$~\citep{warnatz1984}; $A_2= 1.51\times10^{15}$, $n_2=0$, and $E_2=0$~\citep{umist2012};
and $A_{3a}= 7.95\times10^{21}$, $n_{3a}=-1.37$, and $E_{3a}=0$;
and $A_{3b}=1.25\times10^{19}$, $n_{3b}=-0.5$, and $E_{3b}=0$~\citep{prager2007} (units are cm, mol, s, kJ, and K).

The \hyd~and \for~ions are key participants in the chemistry of charged species
because \hyd is the most abundant cation in lean to stoichiometric
hydrocarbon flames~\citep{ouruqpaper2015} and, although only present
in trace amounts, \ce{HCO+} plays a role in the fast proton transfer reaction
(R2) leading to \hyd.
The thermodynamic properties of charged species are taken from the Burcat database~\citep{Burcat2006}. 
Ion transport is modeled according to a mixture-average approach with potentials adequate for charged species~\citep{jieiontransport2015}.
The mobility of the electron is constant and equal to $\mu_e = 0.4$ m$^2$ V$^{-1}$ s$^{-1}$~\citep{fabrizio_etransport2012}.
Einstein's relation is used to obtain diffusion coefficients from mobilities and vice versa.
A complete description of the modeling framework is available in Refs.~\citep{jieiontransport2015,belhi2015}.

It is widely recognized that electrons, ions and neutrals take part in numerous charge transfer reactions,
which give rise to many positive and negative ions in addition to those considered here~\citep{ouruqpaper2015}.
Nonetheless, the total positive charge in a flame
is not affected significantly by the complex, and largely uncertain, network of proton 
transfer reactions, so that the total concentration of cations depends
mostly on the rates of chemi-ionization (R1) and, to a lesser extent, those of recombination (R3)~\citep{ouruqpaper2015}.
Accordingly, and since the concentration of \ce{HCO+} is negligible,
the \hyd~ion is representative of the totality of the positive ions in a flame.
The skeletal ion mechanism used here provides a sensible representation of the negative charges
in the burnt gases downstream of the reaction zone, where electrons abund.
Conversely, negative charges in the unburnt region are mostly anions
ensuing from electrons attaching to neutrals (e.g.~\ce{O2})~\cite{prager2007}.
Since the mobility of electrons is much higher than that of anions, assuming
that current densities associated to negative charges in the unburnt gases
are due to electrons only may lead to errors.
Neglecting anions is not expected to have any adverse effects
in the case of voltage polarities resulting in electrons moving 
from the reaction zone into the burnt region.
The model does not involve impact ionization reactions since the
induced electric field ($E/N$), even at saturation, is considerably
below breakdown limit \citep{speelman2015,yxiV2016,fabrizio_etransport2014}.

\begin{figure}[htbp]
\centering\includegraphics[width=\one]{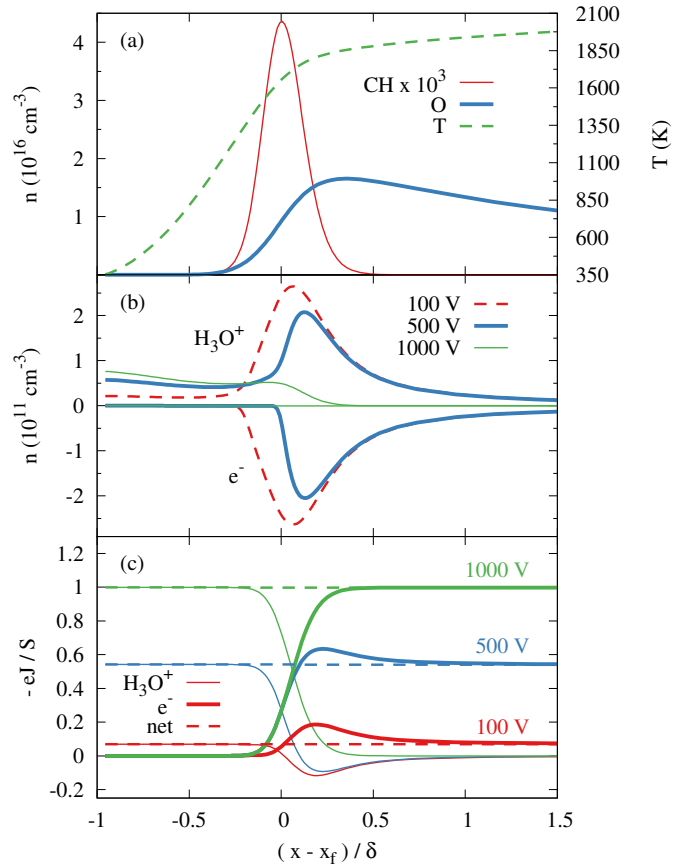}
\caption{Reactive scalars, number density of charges, and current density in a stoichiometric,
burner-stabilized methane/air flame.
The configuration reproduces that in Ref.~\citep{speelman2015}.
(a) Temperature ($T$) and number densities of \ce{CH} and \ce{O};
(b) number densities of \ce{e-} and \ce{H3O+}
for various potential differences: 100, 500, and 1000 V;
(c) charge fluxes (\ce{H3O+}, \ce{e-}, and net)
normalized by the saturation current density $S=0.2$ mA cm$^{-2}$ (Eq.~\eqref{eq:S}).
The saturation voltage is $\approx 800$ V.}
\label{fig:structure}
\end{figure}

\section{Electrical aspects of burner-stabilized flames under applied voltage}\label{sec:aspects}

Ions and electrons are formed in the reaction zone where \ce{CH} and \ce{O}
radicals are abundant (Fig.~\ref{fig:structure}(a)).
In the following analysis, $x_f$  denotes the location of peak \ce{CH} and is taken to represent the location of the
flame. Our simulation shows that $x_f$ coincides approximately with the location of peak heat release rate,
where the ion concentration is maximum \cite{ionflatflame1965}.
The thermal flame thickness is $\delta = (T_b - T_u) / \max\{ dT/dx\}$,
and for the stoichiometric flame, $x_f = 0.41$ mm and $\delta= 0.43$ mm.
Note that $x=0$ corresponds to the left boundary at the surface of the burner. 

\hyd~and \ele~diffuse out of the reaction zone
due to the concurrent effects of the self-sustained electric field and mass diffusion.
In the absence of an applied voltage, the self-sustained electric field inside the reaction zone slows down
the diffusion of the highly mobile electrons and accelerates
that of ions, resulting in the \emph{ambipolar diffusion} process
and a zero net current from the flame.

The saturation phenomenon observed in $\iV$ curves is related to the movement of ions and electrons
induced by the electric field~\citep{electriclaspect1969}.
Below, we discuss the case $\Delta V= V_R - V_L >0$, where the two electrodes
are placed on the left- and right boundaries of the domain, corresponding to $x=0$ at
the burner's surface and $x=L$ on the burnt side of the flame, respectively.
In the case of a sufficiently high (positive) potential difference between the two electrodes,
electrons flow from the reaction zone towards the right electrode, while cations
flow from the reaction zone towards the left electrode.
Figure~\ref{fig:structure}(b) shows that, as the potential difference $\Delta V$ increases,
electrons are removed from the unburnt region of the flame.
Positive ions are abundant in the preheated zone
and some leakage of cations on the right-hand side of the flame persists at all potential differences.
The concentrations of negative and positive charges are asymmetric
around the reaction zone when voltages are applied, reflecting important
differences in the mobilities of electrons and cations to be discussed later.

The interpretation of the effects of applied voltage on the distribution of charged species
and on the current drawn from a flame
are described more appropriately by the number density fluxes of charged species rather than by their charge concentrations.
Recall that the conservation equation for the number density $n_i$ is $dJ_i/dx = \dot{\omega}_i$~\citep{jieiontransport2015,fabrizio_etransport2012},
where $J_i$ is the number flux of species $i$, and $\dot{\omega}_i$ is the source term due to reactions.
By contrast with neutral species, $J_i$ consists of three terms for charged species~\citep{jieiontransport2015,fabrizio_etransport2012}:
\emph{(a)} the convective flux due to the number-averaged bulk velocity,
\emph{(b)} the diffusive flux due to gradients in the number density,
and \emph{(c)} the drift diffusion flux $\pm \mu_i E n_i$, where $\mu_i$
is the mobility of the charged species, $E$ is the
electric field strength, and $+$ ($-$) denotes positive (negative) charges.
Note that in this work we consider the case of singly charged species only.

Figure~\ref{fig:structure}(c) shows the spatial distribution of $eJ^+ =eJ_{\rm H3O+}$ and $eJ^- = eJ_{\rm e-}$
across the flame, where $J^{\pm}$ includes all contributions to the total number density flux,
i.e.~convective, diffusive, and drift terms, and $e$ is the elementary charge.
The quantity $eJ=e(J^+-J^-)$ represents the net flux of charges, or \emph{current density}, and is shown also.
In Fig.~\ref{fig:structure}(c), $eJ<0$, as cations flow towards the left electrode and electrons
flow towards the right electrode in the burnt gases for $\Delta V>0$.

From the data in Fig.~\ref{fig:structure}(c), we note firstly that the current density $eJ$ is 
constant across the flame, regardless of whether a voltage is applied.
Secondly, $eJ$ is negligible when a voltage is not applied, but increases in magnitude as the voltage increases,
in accordance with the $\iV$ curves in Fig.~\ref{fig:ivspeelman}.
Thirdly, $J=J^+$ for $x \lesssim x_f$ and $J=-J^-$ for $x \gtrsim x_f$ for positive voltage
differences ($\Delta V>0$).
In other words, current is due to the flow of cations in the unburnt region and to the flow 
of electrons in the burnt gases.

When voltages reach saturation conditions, the majority of the ions and electrons
generated through chemi-ionization (R1) are removed from the flame, so that
the saturation current density $S$ may be estimated as:
\begin{equation}\label{eq:S}
S = (e/N_A)\int_0^L k_1 [\ce{CH}][\ce{O}] dx,
\end{equation} 
where $N_A$ is the Avogadro number and $[\cdot]$ denotes the number concentration of a species.
It is clear that the saturation current density $S$ is related to the flame
chemistry and reflects the rates of charge production in the reaction zone. 
Note that Eq.~\eqref{eq:S} is an approximation as it neglects recombination processes (R3) 
under the assumption that, at saturation conditions,
ions and electrons are segregated on opposite sides of the flame and are unable to recombine
with any appreciable rate.

In Fig.~\ref{fig:structure}(c), it is shown that $|eJ/S| = 1$ as the voltage increases above the saturation threshold,
confirming the validity of Eq.~\eqref{eq:S}.
According to the detailed model, $S=0.2$ mA cm$^{-2}$ with a saturation voltage $\approx 800$ V for this flame configuration.
We remark that the saturation current density obtained from the detailed model
is about ten times higher than that observed experimentally for the same flame configuration
($i/A\approx0.02$ mA cm$^{-2}$ as shown in Fig.~\ref{fig:ivspeelman}), suggesting that there may be 
significant uncertainties in the predicted concentration of \ce{CH} and/or
in the rate parameters for the chemi-ionization reaction (R1).
The improvement of the rate constants affecting the rates of production of charges in flames
for the purpose of matching $\iV$ curves determined experimentally is beyond the scope of this work
and the rate parameters recommended in the literature are used without modification.

\section{Analytical model for the $\iV$ curve}\label{sec:analytical}

In this Section, we derive an analytical model for distributions of voltage and electric field 
across flames under an applied voltage,
starting from key approximations consistent with the physical processes described in Section~\ref{sec:aspects}.
For the sake of clarity, the relevant equations are derived first for the case $\Delta V>0$,
and second for the case $\Delta V<0$ in Section~\ref{sec:dvnegative}.

Beginning with the drift diffusion flux $J_d^{\pm} = \pm \mu^{\pm} E n^{\pm}$, we derive the expression
$n^{\pm} E = \pm J_d^{\pm} / \mu^{\pm}$, where $n^{\pm}$
are the total number densities of all positive and negative charges
and the mobilities of these charges are denoted by $\mu^+$ and $\mu^-$, respectively.
Upon subtracting the expressions for the total number densities of
positive and negative charges, we obtain
\begin{equation}\label{eq:netflux0}
 (n^+ - n^-)E = J_d^+/\mu^+ + J_d^-/\mu^-.
\end{equation}
\begin{figure}[htbp]
\centering\includegraphics[width=\one]{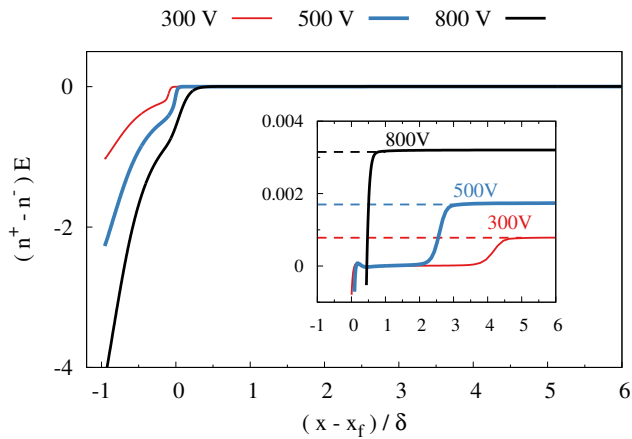}
\caption{$(n^+ - n^-)E$ across the flame for
various values of the applied potential difference $\Delta V$: 300 (solid thin line),
500 (solid thick line), and 800 V (dashed line).
Symbols correspond to the sum of the mobility-weighted
total number density fluxes $J^+/\mu^+ + J^-/\mu^-$:
300 (open squares), 500 (open circles), and 800 V (open triangles).
For the sake of clarity, the inset shows the two quantities over a range of values
close to zero.
The label $A$ corresponds to the unburnt region between the burner
and the reaction zone ($x_f$).
The labels $B_{300}$ and $C_{300}$ correspond to the dead zone region
and the region between the rightmost edge of the dead zone ($x_{d,300}$)
and the right (burnt side) electrode, respectively,
for $\Delta V=300$ V. The current reaches saturation at $800$ V.}
\label{fig:nEJ}
\end{figure}
We note that the left-hand-side of Eq.~\eqref{eq:netflux0}
is the electrohydrodynamic force per unit charge,
which results in what is commonly known as the \emph{ionic wind effect}.
Next, we let $J_d^+/\mu^+ + J_d^-/\mu^- \approx J^+/\mu^+ + J^-/\mu^-$.
This approximation is valid as shown in Fig.~\ref{fig:nEJ}, where
$(n^+ - n^-)E$ and $J^+/\mu^+ + J^-/\mu^-$ are compared.
With this approximation, Eq.~\eqref{eq:netflux0} becomes
\begin{equation}\label{eq:netflux}
 (n^+ - n^-)E = J^+/\mu^+ + J^-/\mu^-,
\end{equation}
highlighting an important relation between the local electric field $E$,
the net charge density $(n^+ - n^-)$, and the sum of the mobility-weighted
total number density fluxes $J^+/\mu^+ + J^-/\mu^-$.

In order to derive a reduced-order model, we seek approximations to the right-hand-side of Eq.~\eqref{eq:netflux} based on
the actual spatial distribution of the fluxes of charges in flames under an applied voltage
as obtained from the numerical solutions to the detailed model (see Section~\ref{sec:aspects}).
We begin by noting that $J = const$ across the flame, so that, for $\Delta V>0$,
positive and negative charge fluxes make up entirely the charge flux upstream and downstream
of the reaction zone, respectively.
This observation suggests that the sum of the mobility-weighted fluxes on the right-hand-side of Eq.~\eqref{eq:netflux}
may be approximated as $J/\mu^+$ for $x \ll x_f$ (unburnt side) and $-J/\mu^-$ for $x \gg x_f$ (burnt side),
where $x_f$ is the location of the reaction zone as defined in Section~\ref{sec:aspects}.

The inset in Fig.~\ref{fig:nEJ} shows the detail of the mobility weighted fluxes in the vicinity
of the reaction zone of the flame ($-1 \le (x-x_f)/\delta \le 10$) for various values of the voltage
difference.
While it is apparent that $(n^+ - n^-)E = -J/\mu^- = const$ for $x \gg x_f$ as expected,
we note that in the downstream of the flame there is a region where $(n^+ - n^-)E \approx 0$.
We shall refer to this region as the \emph{dead zone}, with $x_d$ being
its rightmost boundary identified as the location of maximum slope of the $(n^+ - n^-)E$ profile
downstream of the flame shown in the inset in Fig.~\ref{fig:nEJ}.
The dead zone region for $\Delta V=300$ V
is marked with the label $B_{300}$ and its rightmost edge with $x_{d,300}$.
As shown, a dead zone exists for all voltages below saturation.
At saturation conditions, reached at about $800$ V for this flame configuration,
$x_d \to x_f$ and the dead zone shrinks to a point.

In accordance with the trends detailed above, we formulate the following simplified model for $(n^+ - n^-)E$:
\begin{equation}\label{eq:netflux2}
(n^+-n^-)E = \left\{
\begin{array}{ll}
\phantom{-} J/\mu^+, & x \in [0,x_f) \\
\phantom{-} 0, & x \in [ x_f, x_d] \\
-J/\mu^-, & x \in (x_d,L]
\end{array} \right. ,
\end{equation}
where $\mu^+$ and $\mu^-$ are the mobilities for the positive and negative charged particles
in the unburnt and burnt regions of the flame.
Note that the mobilities $\mu^+$ and $\mu^-$ are treated as constants in Eq.~\eqref{eq:netflux2}.
In reality, while $\mu^- = \mu_e \approx const$ in the burnt gases~\citep{fabrizio_etransport2012},
the cation mobility $\mu^+$ does vary significantly across the preheat zone ahead of the flame.

The differential form of Gauss's law for the electric field implies that $e(n^+-n^-)  = \epsilon_0 dE/dx$,
where $\epsilon_0$ is the vacuum permittivity.
Gauss's law is combined with Eq.~\eqref{eq:netflux2} to obtain a differential equation for $E^2$.
For the sake of clarity, the governing differential equation is derived in nondimensional form using reference
quantities and letting $x=\widetilde{x}L$, $J= \widetilde{J}S/e$, $\mu = \widetilde{\mu} \mu_e$,
$E = \widetilde{E} [8LS/(9\epsilon_0\mu_e)]^{1/2}$, and $V = \widetilde{V}[8L^3S/(9\epsilon_0\mu_e)]^{1/2}$,
where $\widetilde{\cdot}$ denotes a nondimensional
quantity, $S$ is the saturation current density as defined in Eq.~\eqref{eq:S}, and $\mu_e$ is the constant electron mobility.
In nondimensional form, the differential equation for $\widetilde{E}^2$ reads
\begin{equation}\label{eq:Eode}
\frac{4}{9}\f{d\widetilde{E}^2}{d\widetilde{x}}=\left\{
\begin{array}{ll}
\phantom{-} \widetilde{J}/\widetilde{\mu}^+, & \widetilde{x} \in [0,\widetilde{x}_f) \\
\phantom{-} 0, & \widetilde{x} \in [\widetilde{x}_f,\widetilde{x}_d] \\
-\widetilde{J}/\widetilde{\mu}^-, & \widetilde{x} \in (\widetilde{x}_f,1]
\end{array} \right. .
\end{equation}
with solution
\begin{equation}\label{eq:ede1}
E^2(x)=\left\{
\begin{array}{ll}
- (9 J/4\mu^+) (x_f-x) + E^2_f, & x \in (0,x_f] \\
\phantom{-} E^2_f, & x \in [x_f,x_d] \\
- (9 J/4\mu^-) (x-x_d) + E^2_f, & x \in (x_d,1]
\end{array} \right. ,
\end{equation}
where the $\widetilde{\cdot}$ is dropped and all quantities are taken as nondimensional henceforth,
except otherwise stated.
Equation~\eqref{eq:ede1} is general and applicable to any voltage
difference, while the values taken by $x_d$ and $E_f$ qualify the solution to
one of three regimes \--- sub-saturation, saturation, and super-saturation \---
as it will be discussed next.

\subsection{Sub-saturation regime}
\begin{figure}[htbp]
\centering
\centering\includegraphics[width=\one]{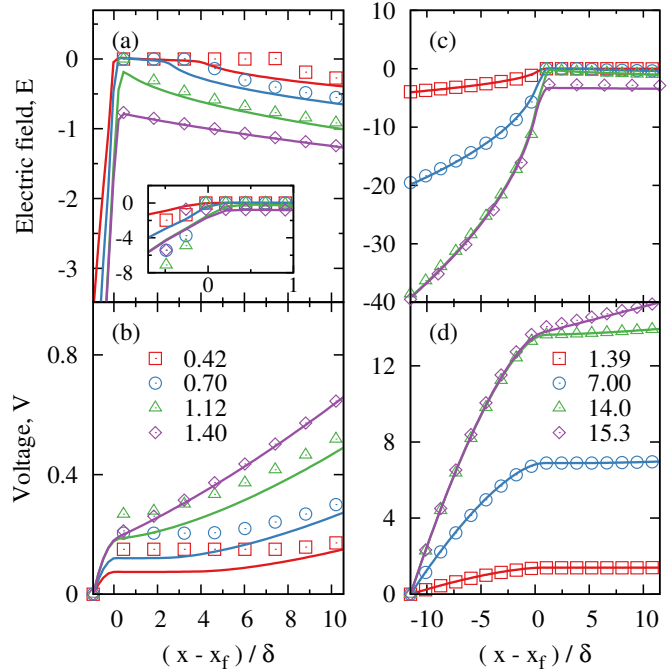}
\caption{Spatial distribution of the electric field $E$ (top row, (a) and (c))
and voltage $V$ (bottom row, (b) and (d)) in the stoichiometric,
methane/air burner-stabilized flame for various values of the applied voltage $\Delta V$
with $x_f=0.04$ (left column, (a) and (b)) and $x_f=0.54$ (right column, (c) and (d)), corresponding
to the reaction zone being close to the burner (left electrode) or further away from it, respectively.
The data from the numerical simulation with the detailed model (solid lines)
and the analytical model (symbols) are shown.
$\Delta V_s=1.12$ and $14.0$ for $x_f=0.04$ and $x_f=0.54$, respectively.}
\label{fig:efV}
\end{figure}

At sub-saturation conditions, $\Delta V < \Delta V_s$ and $I=-J < 1$, where $I$ is the nondimensional current density.
As shown in Fig.~\ref{fig:efV}, at sub-saturation conditions, the electric field inside the dead zone is almost zero
and significantly smaller than the electric field outside it.
Thus, we let $E_f=0$ and Eq.~\eqref{eq:ede1} gives:
\begin{equation}\label{eq:lowef}
E(x)=\left\{
\begin{array}{ll}
-(9I/4\mu^+)^{1/2}(x_f-x)^{1/2}, & x\in[0,x_f)\\
\phantom{-}0, & x\in[x_f,x_d]\\
-(9I/4\mu^-)^{1/2}(x-x_d)^{1/2}, & x\in(x_d ,1]
\end{array} \right. .
\end{equation}
The voltage distribution $V(x)$ is recovered upon integrating Eq.~\eqref{eq:lowef}: 
\begin{equation}\label{eq:lowV}
V(x)=\left\{
\begin{array}{ll}
-(I/\mu^+)^{1/2}(x_f-x)^{3/2} + V_f, & x \in [0,x_f)\\
\phantom{-}V_f, & x\in[x_f,x_d]\\
\phantom{-}(I/\mu^-)^{1/2}(x-x_d)^{3/2} + V_f, & x\in(x_d,1]
\end{array} \right. ,
\end{equation}
where $V_f=V_L + (I/\mu^+)^{1/2}x_f^{3/2}$ is the constant \emph{flame voltage} in the dead zone,
and $V(0)=V_L$, representing the voltage at the left electrode.
Note that $V_f$ increases with increasing current density $I$
and depends on $\mu^+$, i.e.~the mobility of the cations in the pre-heat zone on the unburnt side of the flame.

Application of a voltage $V_R$ at the right electrode gives
\begin{equation}\label{eq:vR}
\Delta V = (I/\mu^-)^{1/2}(1-x_d)^{3/2} + (I/\mu^+)^{1/2} x_f^{3/2}.
\end{equation}
Equation~\eqref{eq:vR} provides a functional relation between the current density $I$ and the applied potential difference $\Delta V$
and is applicable for sub-saturation conditions, thereby describing the $\iV$ curve of a flame
stabilized at a distance $x_f$ from the left electrode.

A closed expression for $x_d$ is required in order for Eq.~\eqref{eq:lowef},~\eqref{eq:lowV}, and~\eqref{eq:vR} to be applicable 
without \emph{a priori} knowledge of the electric structure of a flame/electrode system.
We postulate that the location of the dead zone $x_d$ is a function of
of the nondimensional current denstiy $I$ and the flame location $x_f$ alone,
i.e., $x_d=f(I,x_f)$. The data from the detailed numerical simulations were used to
seek an expression for $x_d=f(I,x_f)$, noting that as $I\to1$, it must follow
that $x_d\to x_f$, as discussed above.

\begin{figure}[htbp]
\centering
\centering\includegraphics[width=\one]{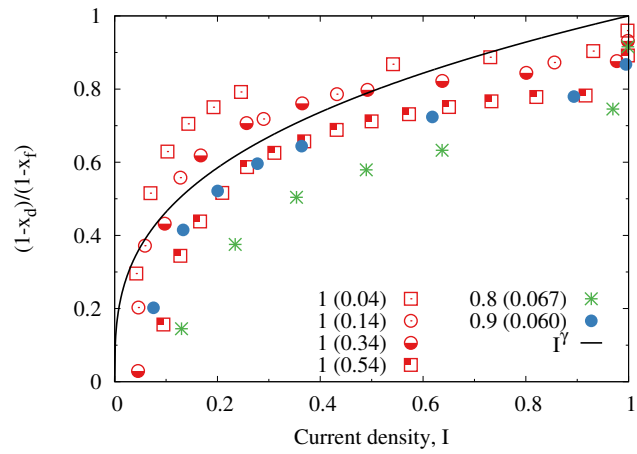}
\caption{$(1-x_d)/(1-x_f)$ versus the current density $I$
for stoichiometric and lean methane/air flames ($\Phi=0.8,0.9$)
obtained from the numerical solution to the detailed model.
For the stoichiometric flame, flames stabilizing at various
distances $x_f = \{0.04,0.14,0.34,0.54\}$ are considered.
For the two additional lean flames, the stabilization is $x_f=0.067$
and $x_f=0.060$ for $\Phi=0.8$ and 0.9, respectively.
The power law $I^{\gamma}$ with $\gamma=1/3$ is shown as a solid line.}
\label{fig:scaling}
\end{figure}

In Fig.~\ref{fig:scaling}, the data from various numerical simulations
are reported in the form $(1-x_d)/(1-x_f)$ versus $I$
for lean and stoichiometric flames with various flame locations $x_f$.
The collapse of the data is reasonable albeit not perfect, suggesting that other parameters may be important,
e.g.~the ratio of the flame thickness to the electrode separation distance, $\delta/L$,
or the flame stoichiometry.
Significantly more work over a broad range of configurations is needed in order to
characterize additional dependencies in a comprehensive fashion.
This analysis is ongoing and will be reported in a future work.
The dependence of $x_d$ on $I$ is described by a power law with exponent $\gamma$:
\begin{equation}\label{eq:scaling}
1-x_d = (1-x_f) I^{\gamma}.
\end{equation}
Based on a least-squares fit as shown in Fig.~\ref{fig:scaling}, we let $\gamma = 1/3$.

Equation~\eqref{eq:scaling} is substituted into Eq.~\eqref{eq:vR} to obtain:
\begin{align}
\label{eq:generalsolution}
\Delta V & = I^{(1+3\gamma)/2} \alpha + I^{1/2}\beta,\\
\label{eq:particular}
\Delta V & = I \alpha + I^{1/2}\beta, 
\end{align}
where $\alpha = ((1-x_f)^3/\mu^-)^{1/2}$ and $\beta = (x_f^3/\mu^+)^{1/2}$,
and $\gamma=1/3$ in Eq.~\eqref{eq:particular}, as suggested by our analysis. 
For a given value of $\Delta V$, $I$ is obtained as a solution to Eq.~\eqref{eq:particular}. 

In Fig.~\ref{fig:efV}, the electric field and voltage implied by
the analytical model above, in which $x_d$ is computed according to Eq.~\eqref{eq:scaling}
with $\gamma=1/3$, are compared to those from the numerical solution from the detailed model,
and are found to be accurate throughout most of the burnt region for all sub-saturation voltages.
Two flame configurations are shown, one with $x_f=0.04$, and the other with $x_f=0.54$, corresponding
to the reaction zone being situated close to the burner (the left electrode) or further away from it, respectively.
The utility of the dead zone concept is apparent for $x_f=0.04$, as a region at constant voltage $V=V_f$
extending downstream of the flame is apparent.
It should be noted that, in the unburnt region, the analytical expression
for the electric field is not quite as accurate for the case $x_f=0.04$ (see inset in Fig.~\ref{fig:efV}(a))
since $\mu^+$ is not constant, as was assumed in the derivation of Eq.~\eqref{eq:lowef}.
This is also evident in Fig.~\ref{fig:nEJ}, where the approximation $J/\mu^+ = const$
is clearly not satisfactory for unburnt gases, where $\mu^+$ varies with temperature.
The treatment of the cation mobility as constant is most likely responsible for
the somewhat inaccurate prediction of the flame voltage $V_f = V_L + (I/\mu^+)^{1/2}x_f^{3/2}$,
which is shown to be overpredicted by the analytical model in Fig.~\ref{fig:efV}(b).

Equation~\eqref{eq:particular} can be used to show the complex dependence of the electrical
response of the flame on the distance of the reaction zone from the electrodes.
The first configuration considered reproduces the experimental setup of Speelman et al.~\citep{speelman2015} and
consists of a flame that stabilizes very close to the burner ($x_f=0.04$) acting as the left electrode.
The second configuration consists of a flame that stabilizes far away from the left electrode ($x_f=0.54$).

Given the nondimensional mobilities of electrons in the burnt region ($\mu^-=1$)
and cations in the unburnt region ($\mu^+ = 1/1300$), we have $\alpha=0.95$ and $\beta=0.19$
for $x_f=0.04$, while $\alpha=0.32$ and $\beta=15$ for $x_f=0.54$.
Thus, for a flame that is very close to the left electrode, $\alpha \gg \beta$,
$\Delta V = I\alpha$, and the current varies linearly with the potential difference.
For a flame that is further away from the left electrode, $\beta \gg \alpha$,
$\Delta V = I^{1/2} \beta$ or $I = \Delta V^2 / \beta^2$,
and the current varies quadratically with the applied potential difference.
Since $\alpha=\beta$ for $x_f \approx 0.08$, the transition from
$I \propto \Delta V$ to $I \propto \Delta V^2$ occurs at relatively
small distances from the left electrode.

\begin{figure}[htbp]
\centering\includegraphics[width=\one]{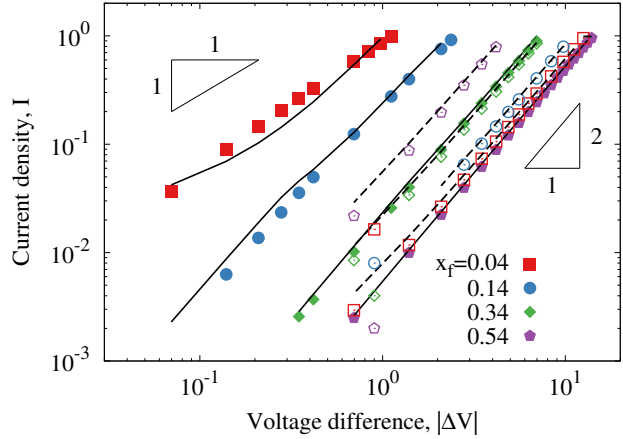}
\caption{$\iV$ curves for various values of $x_f$ from the analytical model (symbols)
and the numerical solution of the detailed model (lines).
Data for $\Delta V >0$ (closed symbols and solid lines)
and $\Delta V<0$ (open symbols and dashed lines) are shown.}
\label{fig:iv}
\end{figure}
Figure~\ref{fig:iv} shows the current $I$ versus the applied potential difference $\Delta V$ for various
values of $x_f$ as obtained from the detailed numerical solutions and from
the analytical model. 
For all values of $x_f$, the analytical model is found to be quite accurate and
the regime transition from $I \propto\Delta V$ to $I \propto\Delta V^2$ is apparent as $x_f$ increases.
For small $x_f$, the limit $I \propto\Delta V$ is recovered, although the exponent is only approximately equal to unity
since the power law behavior for $\alpha \gg \beta$ depends on the value of $\gamma$
in Eq.~\eqref{eq:generalsolution}, which is $\approx 1/3$ (Eq.~\eqref{eq:particular})
and clearly displays residual dependence on flame parameters not captured by Eq.~\eqref{eq:scaling}.
Conversely, the limit $\Delta V \propto I^2$ for $\beta \gg \alpha$ is exact
since it does not depend on the model closure for $x_d$.

\begin{figure}[htbp]
\centering\includegraphics[width=\one]{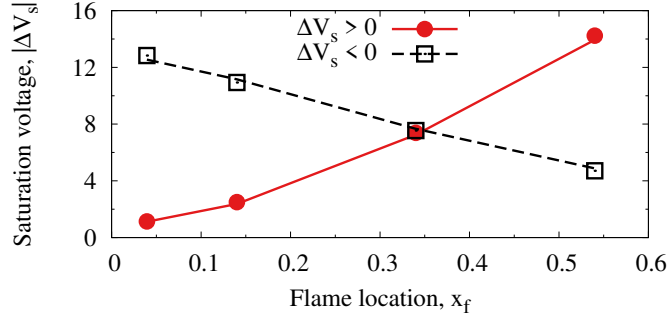}
\caption{Saturation voltage $|\Delta V_s|$ as a function of the flame location $x_f$:
detailed model (symbols) and analytical model (lines) as in Eq.~\eqref{eq:vr1}
for positive voltages and Eq.~\eqref{eq:vR2} with $x_d=x_f$ and $\mu^+=1/180$ for negative voltages.}
\label{fig:deltaV}
\end{figure}

\subsection{Saturation conditions}

At saturation conditions, $I=1$, $E_f=0$,
and $x_d = x_f$ (see Fig.~\ref{fig:nEJ} and Fig.~\ref{fig:efV}), and the \emph{saturation voltage}
$\Delta V_s=V_{R,s}-V_L > 0$ is defined as
\begin{equation}\label{eq:vr1}
\Delta V_s = ((1-x_f)^3/\mu^-)^{1/2} + (x_f^3/\mu^+)^{1/2}.
\end{equation}
The flame voltage is equal to $V_f = (1/\mu^+)^{1/2}x_f^{3/2} + V_L$.
The saturation voltage from the numerical solution of the detailed
model is compared to that implied by Eq.~\eqref{eq:vr1}
in Fig.~\ref{fig:deltaV} and found to be in excellent agreement.
A similar expression was also derived
and validated via experiments on burning droplets in a vertical DC field~\citep{samedrivation2011}.

A significant implication of Eq.~\eqref{eq:vr1} is that, for a given fuel and equivalence ratio, the nondimensional
potential difference required to reach saturation depends both 
on the flame location relative to the two electrodes, and on the nondimensional mobilities of the positive and negative charges.
For example, for $x_f=0.04$ and $0.54$, we have \mbox{$\Delta V^*_s \approx 0.8$} and $\approx 10$ kV, respectively,
where $(\cdot)^*$ denotes a dimensional quantity.
Therefore, we conclude that flames that stabilize away from the left electrode require significantly
higher potential differences to reach saturation conditions. 
Given the reference voltage employed for the nondimensionalization,
$\Delta V^*_s \propto S^{1/2} L^{3/2} \mu_e^{-1/2}$ and
the saturation voltage increases with increasing propensity of the fuel
and/or mixture to form charges. 

\subsection{Super-saturation regime}\label{sec:supersaturation}
As $\Delta V$ exceeds $\Delta V_s$ (Eq.~\eqref{eq:vr1}), super-saturation
conditions are achieved and $I=1$. 
As shown in Fig.~\ref{fig:efV}, super-saturation conditions for the
case $\Delta V>0$ are characterized by an electric field dropping below zero
everywhere across the flame.
As the potential difference increases further, so does $|E_f|$. 

Beginning with Eq.~\eqref{eq:ede1}, we let $J=-1$, which is compatible
with the current density being saturated, and integrate the equation $dV/dx = -E(x)$
with $E_f \neq 0$.
A piecewise solution for the voltage $V(x)$ in the unburnt
($0\le x_f$) and burnt ($x_f < x \le 1$) regions is obtained.
The solution features two integration constants, one for each piece,
and the unknown electric field strength $E_f$.
If we apply the boundary conditions $V(0)=V_L$, $V(1)=V_R$ and the continuity
of the voltage at $x_f$, a non-linear algebraic equation for $E_f$ is derived:
$3\Delta V / 2 = [ ( E_f^2 - a x_f)^{3/2} - E_f^3 ]/a
- [ ( E_f^2 - b(1-x_f))^{3/2} - E_f^3 ]/b$,
where $\Delta V > 0$, $a=9/(4\mu^+)$, and $b=9/(4\mu^-)$.
For a given $\Delta V$, this equation is solved for $E_f$
and the electric field and voltage fields obtained. 
Results are presented in Fig.~\ref{fig:efV},
showing excellent agreement between model and detailed numerical simulations
for $\Delta V <0$.

\subsection{Negative potential difference, $\Delta V<0$}\label{sec:dvnegative}
When $\Delta V<0$, electrons move towards the left electrode (unburnt) and
cations move in the opposite direction.
A dead zone is formed between the left electrode and the flame,
and the solution to the voltage at sub-saturation conditions is:
\begin{equation}\label{eq:lowV2}
V(x)=\left\{
\begin{array}{ll}
\phantom{-}(I/\mu^-)^{1/2}(x_d-x)^{3/2} + V_f, & x \in [0,x_d)\\
\phantom{-}V_f, & x\in[x_d,x_f]\\
-(I/\mu^+)^{1/2}(x-x_f)^{3/2} + V_f, & x\in(x_f,1]
\end{array} \right. ,
\end{equation}
leading to
\begin{equation}\label{eq:vR2}
\Delta V = (I/\mu^+)^{1/2}(1-x_f)^{3/2} + (I/\mu^-)^{1/2} x_d^{3/2},
\end{equation}
We let $\mu^+=1/180$ for cations in the burnt region, while $\mu^{-}=1$.
Unless the flame is situated very close to the right electrode ($x_f\to1$),
the inequality \mbox{$(1-x_f)^{3/2}/(\mu^+)^{1/2}\gg x_d^{3/2}/(\mu^-)^{1/2}$} is valid
and Eq.~\eqref{eq:vR2} simplifies to $\Delta V = (I/\mu^+)^{1/2}(1-x_f)^{3/2}$.
This result is consistent with the data in Fig.~\ref{fig:iv},
where the dependence $I \propto \Delta V^2$ is shown for $\Delta V<0$.

In Fig.~\ref{fig:iv}, it is also apparent that, if the flame is sufficiently close to
the left electrode (e.g.~$x_f=0.04$), the current density $I$ for positive
voltages is much higher than that for negative voltages of equal absolute value.
This behavior is in qualitative agreement with the experimental data reported in Fig.~\ref{fig:ivspeelman},
which features a premixed flame stabilized very close to the burner/left-electrode. 

We caution that the model for $\Delta V < 0$ with $\mu^-=1$ is strictly
valid if electrons contribute the most to the negative charge current density
in the unburnt gases and anions are neglected.
In principle, one could let $\mu^-=1/1300 \ll 1$ in Eq.~\eqref{eq:vR2} in order
to account for the lower mobility of anions. 
Validation of this simple remedy requires the numerical solution
to detailed ion transport and chemistry models
with anions and related electron attachment reactions.
This analysis is ongoing and will be presented at a later time.

\section{Conclusions}\label{sec:conclusions}

We have developed an analytical model to describe the electronic
structure of a burner-stabilized premixed flame subject to an applied voltage.
The proposed model is comprehensive in that it captures all conditions and regimes
\--- saturation, sub-, and super-saturation \--- depending on the applied voltage.
In order to explain the observed voltage distribution for sub-saturation conditions,
we introduce the concept of a \emph{dead zone},
in which the electric field vanishes and the voltage is constant.

The flame location relative to the electrode in the unburnt region
was found to have a significant effect on the saturation
voltage and produced both linear and quadratic scaling relations for the $\iV$ curve response.
Our results were found to be in good agreement with detailed numerical simulation data.
Since the reduced model relies on the existence of a thin layer where charges are produced
(i.e.~the reaction zone), the model is not limited to premixed flames
and application to other types of flames and configurations is possible,
provided that the electrode gap is large compared to the flame thickness.

\section*{Acknowledgments}
The research reported in this publication was supported by King Abdullah University of
Science and Technology (KAUST).


\bibliographystyle{elsarticle-num-CNF}
\bibliography{paper}

\begin{thebibliography}{20}
\expandafter\ifx\csname natexlab\endcsname\relax\def\natexlab#1{#1}\fi
\providecommand{\bibinfo}[2]{#2}

\bibitem[{Lawton and Weinberg(1969)}]{electriclaspect1969}
\bibinfo{author}{J.~Lawton}, \bibinfo{author}{F.~J. Weinberg},
  \bibinfo{title}{Electrical aspects of combustion}, Clarendon Press,
  \bibinfo{address}{Ely House, London}, \bibinfo{year}{1969}.

\bibitem[{Goodings et~al.(2001)Goodings, Guo, Hayhurst, and
  Taylor}]{Goodings2001}
\bibinfo{author}{J.~Goodings}, \bibinfo{author}{J.~Guo},
  \bibinfo{author}{A.~Hayhurst}, \bibinfo{author}{S.~Taylor},
  \bibinfo{journal}{Int. J. Mass.Spectrom.} \bibinfo{volume}{206}
  (\bibinfo{year}{2001}) \bibinfo{pages}{137--151}.

\bibitem[{Kamani and Dunn-Rankin(2015)}]{ddrankiniV2015}
\bibinfo{author}{S.~Kamani}, \bibinfo{author}{D.~Dunn-Rankin},
  \bibinfo{journal}{Combust. Flame} \bibinfo{volume}{162}
  (\bibinfo{year}{2015}) \bibinfo{pages}{2865--2872}.

\bibitem[{Speelman et~al.(2015)Speelman, {de Goey}, and {van
  Oijen}}]{speelman2015}
\bibinfo{author}{N.~Speelman}, \bibinfo{author}{L.~P.~H. {de Goey}},
  \bibinfo{author}{J.~A. {van Oijen}}, \bibinfo{journal}{Combust. Theory Model}
  \bibinfo{volume}{19} (\bibinfo{year}{2015}) \bibinfo{pages}{159--187}.

\bibitem[{Imamura et~al.(2011)Imamura, Chen, Nishida, Yamashita, Tsue, and
  Kono}]{samedrivation2011}
\bibinfo{author}{O.~Imamura}, \bibinfo{author}{B.~Chen},
  \bibinfo{author}{S.~Nishida}, \bibinfo{author}{K.~Yamashita},
  \bibinfo{author}{M.~Tsue}, \bibinfo{author}{M.~Kono}, \bibinfo{journal}{Proc.
  Combust. Inst.} \bibinfo{volume}{33} (\bibinfo{year}{2011})
  \bibinfo{pages}{2005--2011}.

\bibitem[{Xiong et~al.(2016)Xiong, Park, Lee, Chung, and Cha}]{yxiV2016}
\bibinfo{author}{Y.~Xiong}, \bibinfo{author}{D.~G. Park},
  \bibinfo{author}{B.~J. Lee}, \bibinfo{author}{S.~H. Chung},
  \bibinfo{author}{M.~S. Cha}, \bibinfo{journal}{Combust. Flame}
  \bibinfo{volume}{163} (\bibinfo{year}{2016}) \bibinfo{pages}{317--325}.

\bibitem[{Han et~al.(2015)Han, Belhi, Bisetti, and
  Sarathy}]{jieiontransport2015}
\bibinfo{author}{J.~Han}, \bibinfo{author}{M.~Belhi},
  \bibinfo{author}{F.~Bisetti}, \bibinfo{author}{S.~M. Sarathy},
  \bibinfo{journal}{Combust. Theory Model} \bibinfo{volume}{00}
  (\bibinfo{year}{2015}) \bibinfo{pages}{1--29}. \bibinfo{note}{Available at
  \url{http://dx.doi.org/10.1080/13647830.2015.1090018}}.

\bibitem[{Kee et~al.(1985)Kee, Grcar, Smooke, Miller, and
  Meeks}]{premix-chemkin}
\bibinfo{author}{R.~J. Kee}, \bibinfo{author}{J.~F. Grcar},
  \bibinfo{author}{M.~D. Smooke}, \bibinfo{author}{J.~A. Miller},
  \bibinfo{author}{E.~Meeks}, \bibinfo{title}{A Fortran Program for Modeling
  Steady Laminar One-Dimensional Premixed Flames, Report No.~SAND85-8240,
  Sandia National Laboratories}, \bibinfo{type}{Report}, \bibinfo{year}{1985}.

\bibitem[{Belhi et~al.(2013)Belhi, Domingo, and Vervisch}]{DCACBelhi2013}
\bibinfo{author}{M.~Belhi}, \bibinfo{author}{P.~Domingo},
  \bibinfo{author}{P.~Vervisch}, \bibinfo{journal}{Combust. Theory Model}
  \bibinfo{volume}{17} (\bibinfo{year}{2013}) \bibinfo{pages}{749--787}.

\bibitem[{Smooke(1991)}]{skelGRI}
\bibinfo{author}{M.~D. Smooke}, in: \bibinfo{editor}{M.~D. Smooke} (Ed.),
  \bibinfo{booktitle}{Reduced Kinetic Mechanisms and Asymptotic Approximations
  for Methane-Air Flames, Lecture Notes in Physics, Vol.~384}, Springer,
  \bibinfo{address}{Berlin}, \bibinfo{year}{1991}, pp. \bibinfo{pages}{1--28}.

\bibitem[{Smith et~al.(1999)Smith, Golden, Frenklach, Moriarty, Eiteneer,
  Goldenberg, Bowman, Hanson, Song, Jr., Lissianski, and Qin}]{grimech30}
\bibinfo{author}{G.~P. Smith}, \bibinfo{author}{D.~M. Golden},
  \bibinfo{author}{M.~Frenklach}, \bibinfo{author}{N.~W. Moriarty},
  \bibinfo{author}{B.~Eiteneer}, \bibinfo{author}{M.~Goldenberg},
  \bibinfo{author}{C.~T. Bowman}, \bibinfo{author}{R.~K. Hanson},
  \bibinfo{author}{S.~Song}, \bibinfo{author}{W.~C.~G. Jr.},
  \bibinfo{author}{V.~V. Lissianski}, \bibinfo{author}{Z.~Qin},
  \bibinfo{year}{1999}.
  \bibinfo{note}{\url{http://www.me.berkeley.edu/gri_mech}}.

\bibitem[{Warnatz(1984)}]{warnatz1984}
\bibinfo{author}{J.~Warnatz}, in: \bibinfo{editor}{W.~C. Gardiner~Jr} (Ed.),
  \bibinfo{booktitle}{Combustion Chemistry}, Springer US, \bibinfo{year}{1984},
  pp. \bibinfo{pages}{197--360}.

\bibitem[{McElroy et~al.(2013)McElroy, Walsh, Markwick, Cordiner, Smith, and
  Millar}]{umist2012}
\bibinfo{author}{D.~McElroy}, \bibinfo{author}{C.~Walsh},
  \bibinfo{author}{A.~J. Markwick}, \bibinfo{author}{M.~A. Cordiner},
  \bibinfo{author}{K.~Smith}, \bibinfo{author}{T.~J. Millar},
  \bibinfo{journal}{Astron. Astrophys.} \bibinfo{volume}{550}
  (\bibinfo{year}{2013}) \bibinfo{pages}{A36}. \bibinfo{note}{Available at
  {\url{http://udfa.ajmarkwick.net}}}.

\bibitem[{Prager et~al.(2007)Prager, Riedel, and Warnatz}]{prager2007}
\bibinfo{author}{J.~Prager}, \bibinfo{author}{U.~Riedel},
  \bibinfo{author}{J.~Warnatz}, \bibinfo{journal}{Proc. Combust. Inst.}
  \bibinfo{volume}{31} (\bibinfo{year}{2007}) \bibinfo{pages}{1129--1137}.

\bibitem[{Kim et~al.(2015)Kim, Rizzi, Cheng, Han, Bisetti, and
  Knio}]{ouruqpaper2015}
\bibinfo{author}{D.~Kim}, \bibinfo{author}{F.~Rizzi},
  \bibinfo{author}{K.~Cheng}, \bibinfo{author}{J.~Han},
  \bibinfo{author}{F.~Bisetti}, \bibinfo{author}{O.~Knio},
  \bibinfo{journal}{Combust. Flame} \bibinfo{volume}{162}
  (\bibinfo{year}{2015}) \bibinfo{pages}{2904--2915}.

\bibitem[{Burcat(2006)}]{Burcat2006}
\bibinfo{author}{A.~Burcat}, \bibinfo{title}{Ideal gas thermodynamic data in
  polynomial form for combustion and air pollution use}, \bibinfo{year}{2006}.
  \bibinfo{note}{\url{http://garfield.chem.elte.hu/Burcat/burcat.html}}.

\bibitem[{Bisetti and Morsli(2012)}]{fabrizio_etransport2012}
\bibinfo{author}{F.~Bisetti}, \bibinfo{author}{M.~E. Morsli},
  \bibinfo{journal}{Combust. Flame} \bibinfo{volume}{159}
  (\bibinfo{year}{2012}) \bibinfo{pages}{3518--3521}.

\bibitem[{Belhi et~al.(2016)Belhi, Han, Bisetti, Farooq, Sarathy, and
  Im}]{belhi2015}
\bibinfo{author}{M.~Belhi}, \bibinfo{author}{J.~Han},
  \bibinfo{author}{F.~Bisetti}, \bibinfo{author}{A.~Farooq},
  \bibinfo{author}{M.~Sarathy}, \bibinfo{author}{H.~Im},
  \bibinfo{journal}{Combust. Flame} \bibinfo{volume}{1} (\bibinfo{year}{2016})
  \bibinfo{pages}{1--1}. \bibinfo{note}{In preparation}.

\bibitem[{Bisetti and Morsli(2014)}]{fabrizio_etransport2014}
\bibinfo{author}{F.~Bisetti}, \bibinfo{author}{M.~E. Morsli},
  \bibinfo{journal}{Combust. Theory Model} \bibinfo{volume}{18}
  (\bibinfo{year}{2014}) \bibinfo{pages}{148--184}.

\bibitem[{Wortberg(1965)}]{ionflatflame1965}
\bibinfo{author}{G.~Wortberg}, \bibinfo{journal}{Symp. (Int.) Combust.}
  \bibinfo{volume}{10} (\bibinfo{year}{1965}) \bibinfo{pages}{651--655}.

\end{thebibliography}

\end{document}